\documentclass[prl,twocolumn,floatfix,preprintnumbers,amsmath,amssymb,superscriptaddress]{revtex4}

\usepackage{float}
\usepackage{graphicx}
\usepackage{dcolumn}
\usepackage{bm}
\usepackage{color}
\usepackage{xspace}
\usepackage{amsmath}
\usepackage{soul}

\usepackage{textcomp} 

\graphicspath{{./}{figure/}}
\usepackage[colorlinks,plainpages=false,linkcolor=blue,urlcolor=blue,citecolor=blue,pdfpagemode=UseNone,pdfstartview=FitBH]{hyperref}
\usepackage{dcolumn,graphicx,color,booktabs,microtype,afterpage}
\usepackage[charter,greekuppercase=italicized]{mathdesign}

\begin{document}

\title{Magnetic correlations in the semimetallic hyper-kagome iridate Na$_3$Ir$_3$O$_8$}

\author{G.\ Simutis}
\email{gediminas.simutis@universite-paris-saclay.fr}
\affiliation{Laboratoire de Physique des Solides, Paris-Saclay University and CNRS, France}

\author{T. Takayama}
\affiliation{Institute for Functional Matter and Quantum Technologies, University of Stuttgart, Germany}
\affiliation{Max Planck Institute for Solid State Research Stuttgart, Germany}

\author{Q. Barth\'elemy}
\affiliation{Laboratoire de Physique des Solides, Paris-Saclay University and CNRS, France}

\author{F. Bert}
\affiliation{Laboratoire de Physique des Solides, Paris-Saclay University and CNRS, France}

\author{H. Takagi}
\affiliation{Institute for Functional Matter and Quantum Technologies, University of Stuttgart, Germany}
\affiliation{Max Planck Institute for Solid State Research Stuttgart, Germany}

\author{P. Mendels}
\affiliation{Laboratoire de Physique des Solides, Paris-Saclay University and CNRS, France}

\date{\today}

\begin{abstract}

We present a microscopic study of a doped quantum spin liquid candidate, the hyperkagome Na$_3$Ir$_3$O$_8$ compound by using $^{23}$Na NMR. We determine the intrinsic behavior of the uniform \textbf{q} $ = 0$ susceptibility via shift measurements and the dynamical response by probing the spin-lattice relaxation rate. Throughout the studied temperature range, the susceptibility is consistent with a semimetal behavior, though with electronic bands substantially modified by correlations. Remarkably, the antiferromagnetic fluctuations present in the insulating parent compound Na$_4$Ir$_3$O$_8$ survive in the studied compound. The spin dynamics are consistent with 120$^o$ excitations modes displaying short-range correlations.
 \end{abstract}

\maketitle{}

Geometrical frustration in insulating spin systems has long been recognized as a way to access novel quantum states such as spin liquids \cite{Balents2010}. Following the proposal of Anderson that a spin liquid can act as a substrate for superconductivity, the holy grail of the field of frustrated magnetism is the possibility to dope spin liquids, with a promise of emergent new strongly correlated metallic states and possibly even superconductivity \cite{Anderson1987, Lee2006, Mazin2014,Broholm2020}. However, despite many attempts it has remained elusive. In particular, efforts to follow up the ab-initio predictions of doped herbertsmithite \cite{Mazin2014, Kelly2016, Puphal2019} have revealed that doping Zn-Cu hydroxyl halides leaves the electrons localized, possibly due to the formation of polaronic states together with lattice displacements \cite{Liu2018}. Different model systems are therefore needed to tackle this challenge.

The discovery of spin-orbit coupling assisted Mott insulators in iridates has further opened many new possibilities in the study of magnetism \cite{Kim2008}. In addition to their appeal in the context of metal-insulator transitions, they can harbor novel J$_{\mathrm{eff}}$ = 1/2 pseudo-spin states and even provide new forms of frustration, dependent on the bond orientation \cite{Jackeli2009,Kitaev2006,Takagi2019,Trebst2017,Singh2012,Dey2012,Takayama2015,Modic2014}. Since in many cases the electronic gap is of the order of a few hundred meV \cite{Comin2012,Gretarsson2013,Norman2010,He2015}, strong changes of the accessible states can be expected with minor modifications in the compounds.

\begin{figure}
\centering
\includegraphics[width={\columnwidth}]{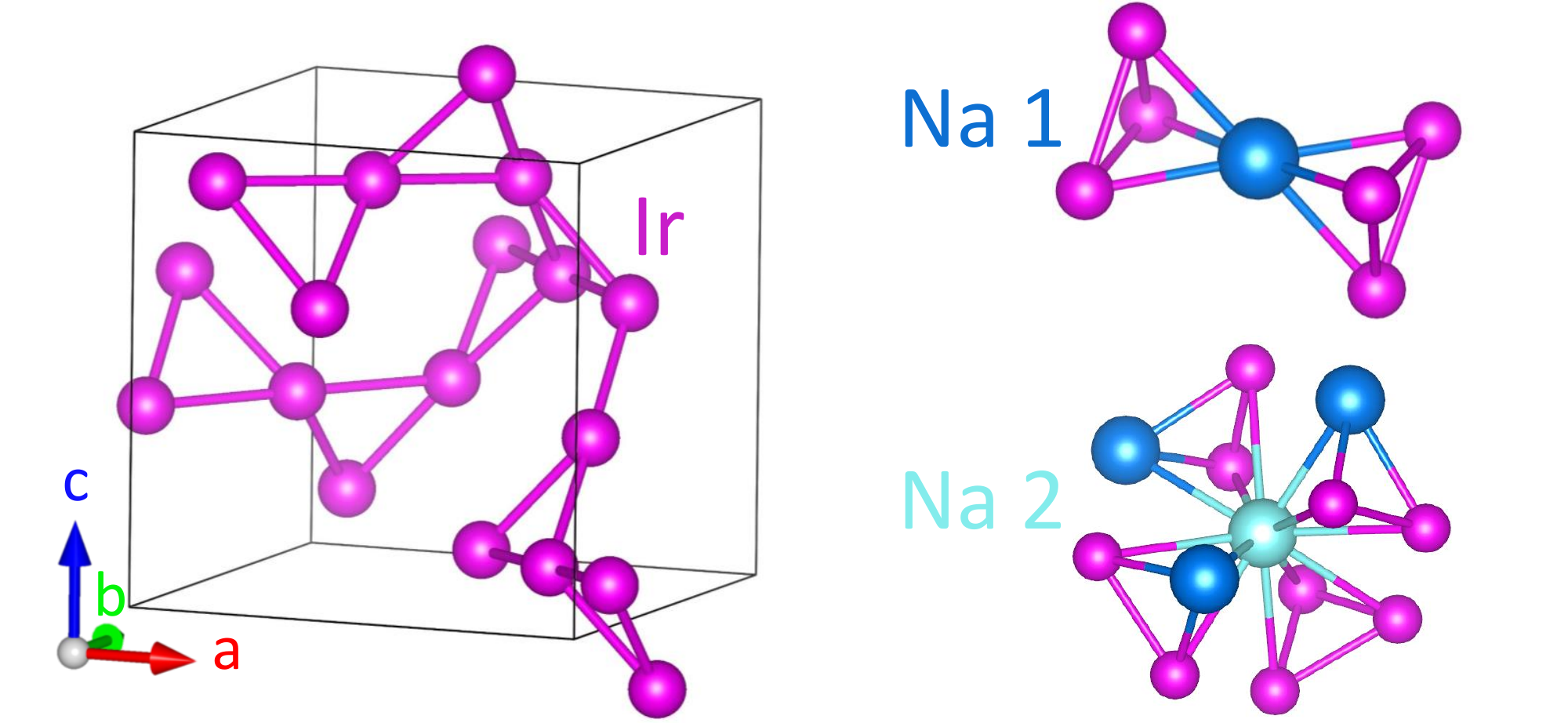}\\
\caption{\label{fig:HKGM}%
The left panel shows the hyperkagome structure of the Iridium ions, present in both Na$_4$Ir$_3$O$_8$ as well as Na$_3$Ir$_3$O$_8$. The oxygen and sodium atoms are removed for clarity. The right panel displays the two different Na environments. The Na 1 nucleus is in between two triangles of Ir with distances to the Ir ions of about $\approx$ 3.3 \AA ~and a total of 6 neighbors. The Na 2 nucleus is connected to 9 nearest Ir neighbors. Three of them are connected as a triangle, similar to the Na 1 case, whereas the three pairs from the remaining 6 Ir atoms form triangles with another Na atom. The distances between the Na 2 and the Ir atoms are $\approx$ 3.7 \AA.
}
\end{figure}

A particularly interesting case is the unique hyperkagome lattice in the form of Na$_4$Ir$_3$O$_8$,\cite{Okamoto2007,Chen2008,Lawler2008,Chen2013} which is an insulator and despite a large effective exchange interaction of J $\approx$ 300 K, only exhibits spin freezing at $\approx$ 7 K \cite{Dally2014,Shockley2015}. The geometric frustration is evident from the arrangement of the iridium ions in the crystal structure, where the triangular motives share the Ir corners in a three-dimensional network as shown in Figure~\ref{fig:HKGM}.

Remarkably, this system can be modified by removing 1/4 of the sodium atoms, which results in Na$_3$Ir$_3$O$_8$. Structurally it maintains the same space group P4$_1$32 (213), but some of the sodium atoms are rearranged: instead of the partially occupied Wyckoff 4a and 12d sites there is a fully occupied 8c site (see a more detailed description and an illustration in \cite{SI2020}). Importantly, the underlying hyperkagome arrangement of iridium ions is preserved \cite{Takayama2014}. The iridium valence in this system is 4.33+ and not 4+ like in the parent compound. In terms of charge balance, it can be considered as a 1/3rd hole doped hyper-kagome spin lattice and indeed the measured resistivity in the Na-deficient system exhibits metallic behavior\cite{Takayama2014,Balodhi2015}. The situation is in fact more complex, calling for an interplay between inter-site hopping, Coulomb repulsion, crystal field splitting and spin-orbit coupling, with a resulting state that appears to be a semi-metal based on Hall measurements and DFT calculations \cite{Takayama2014}. Although the initial expectation has been that application of pressure should lead to a metallic state \cite{Takayama2014}, recent studies showed that instead, a semimetal-to-insulator transition takes place \cite{Sun2018}, further indicating that the system is on the verge of a metal to insulator transition.

In order to understand the complex interplay between the charges and magnetic moments in this material and doped spin liquids in general, two key questions have to be tackled: i) what are the intrinsic magnetic properties of such a system and ii) do the magnetic correlations of a quantum spin liquid persist in a doped material and what role do they play in Na$_3$Ir$_3$O$_8$? In the present study we set out to answer these questions.

It is extremely challenging to access the fundamental properties of the spin correlations in these systems. Since only a small fraction of the electrons close to the Fermi level contributes to the magnetic response, the intrinsic susceptibility is small and the bulk response is dominated by the small number of impurities \cite{Takayama2014}. Additionally, neutron absorption by iridium and a lack of large samples prevent neutron spectroscopy. In this respect, Nuclear Magnetic Resonance (NMR) offers a unique way to investigate both the intrinsic susceptibility through the spectral measurements as a function of temperature as well as access the spin dynamics through relaxation measurements.

In this Letter, we discover that both the static and the dynamic properties indicate the presence of correlations in this system. Most importantly, we determine the nature of these correlations by comparing the two $^{\mathrm{23}}$Na nuclei located at different crystallographic sites. We find that antiferromagnetic correlations of the insulating spin liquid parent compound persist in the semimetallic phase.

\begin{figure}
\centering
\includegraphics[width={\columnwidth}]{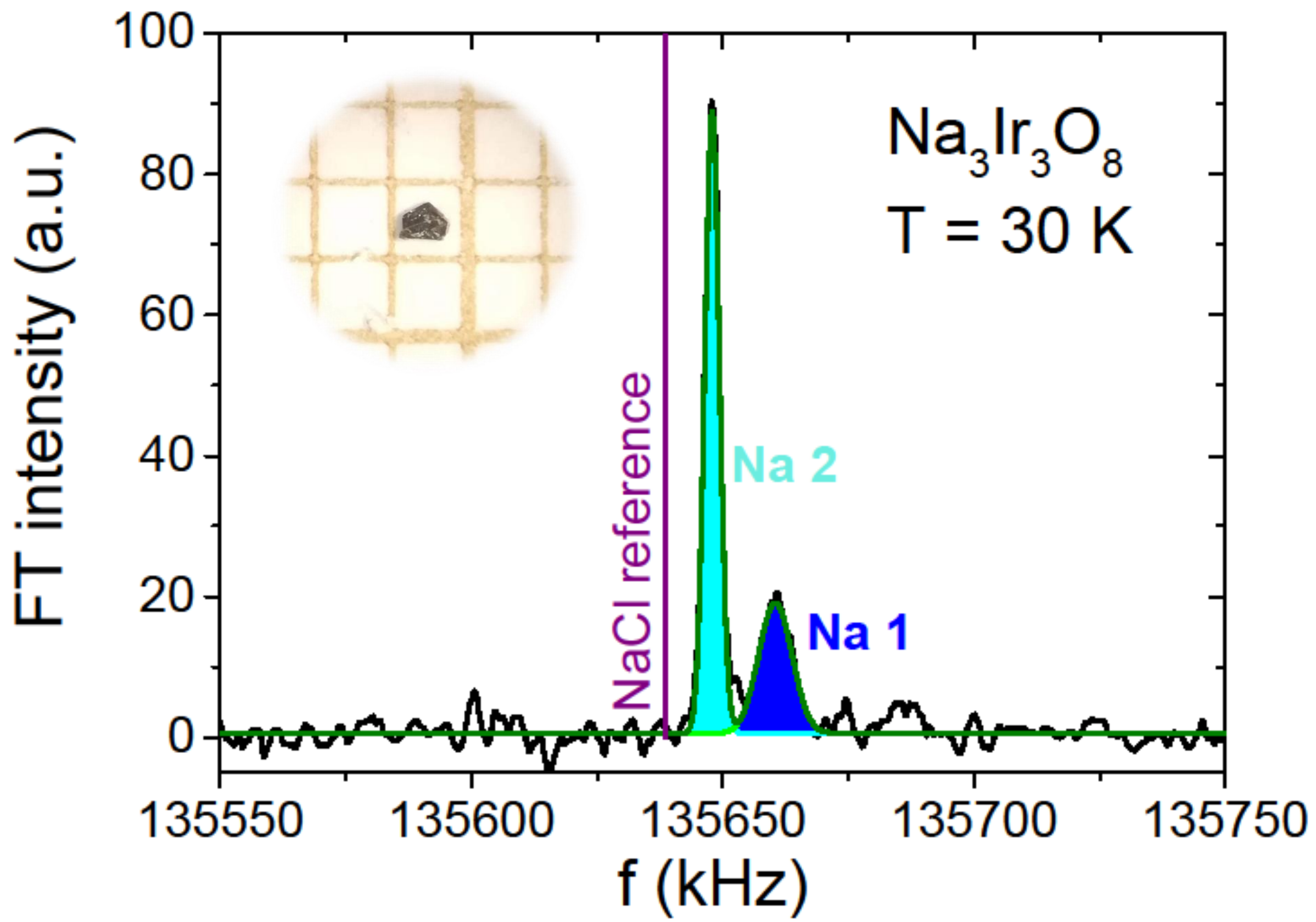}\\
\caption{\label{fig:spectrum}%
$^{23}$Na spectrum taken at a nominal field of 12 T. The two peaks correspond to the two sodium sites in the crystal. The inset shows the sample used in the study.
}
\end{figure}

The experiments were performed on a small sub-milimeter single-crystal sample of Na$_3$Ir$_3$O$_8$ (inset of Fig.~\ref{fig:spectrum}) from the same growth batch as used in Ref.~\cite{Takayama2014}. We used a cryo-free 14 T sweep magnet and a flow cryostat. Additional measurements were performed down to 1.5 K using a cold-bore magnet to exclude any possible spin freezing. All the measurements were obtained using a $\pi/2$-$\pi$ spin echo sequence. For the determination of the spin-lattice relaxation rate, the same experiments were performed with a varying repetition rate to obtain the intensity as a function of the rate. All of the measurements were performed at an applied field yielding the resonance frequency 135638.5 kHz for the $^{23}$Na in NaCl, with a calibration run measuring the spectrum from copper in the coil taken for every measurement, to account for any drift in the field.

The $^{23}$Na nucleus has been shown to be an efficient probe of the intrinsic hyper-kagome physics in the Na$_4$Ir$_3$O$_8$ \cite{Shockley2015} since it has a large gyromagnetic ratio with a 100 \% abundance and is well-coupled to the iridium ions. In Na$_3$Ir$_3$O$_8$ the signal is even cleaner with only two fully-occupied Na sites instead of three. The two inequivalent sodium sites in the structure (Fig.~\ref{fig:HKGM}) lead to two distinct signals as seen in Fig.~\ref{fig:spectrum}. The Na 1 and Na 2 nuclei are positioned at respectively $\approx$ 3.3 \AA ~and $\approx$ 3.7 \AA ~away from the Ir, suggesting that the Na 1 is coupled more strongly and hence shifted more, corresponding to the right-hand-side peak in the spectrum. The ratio of $\approx$ 0.5 between the integrated intensities of the two peaks corresponds to the ratio of the two sites in the structure and confirms this assignment. Since the sodium ions are not located in cubic sites and have a non-zero quadrupolar moment, it would be expected that $^{23}$Na spectra display both a peak from the central transition (1/2 to -1/2) as well as quadrupolar satellites (3/2 to 1/2 and -1/2 to -3/2). However, no satellites were observed in our experiments, likely because of the substantial mosaicity \cite{Takayama2014} which leads to a distribution of the electric field gradients at different sites.

\begin{figure}
\centering
\includegraphics[width={\columnwidth}]{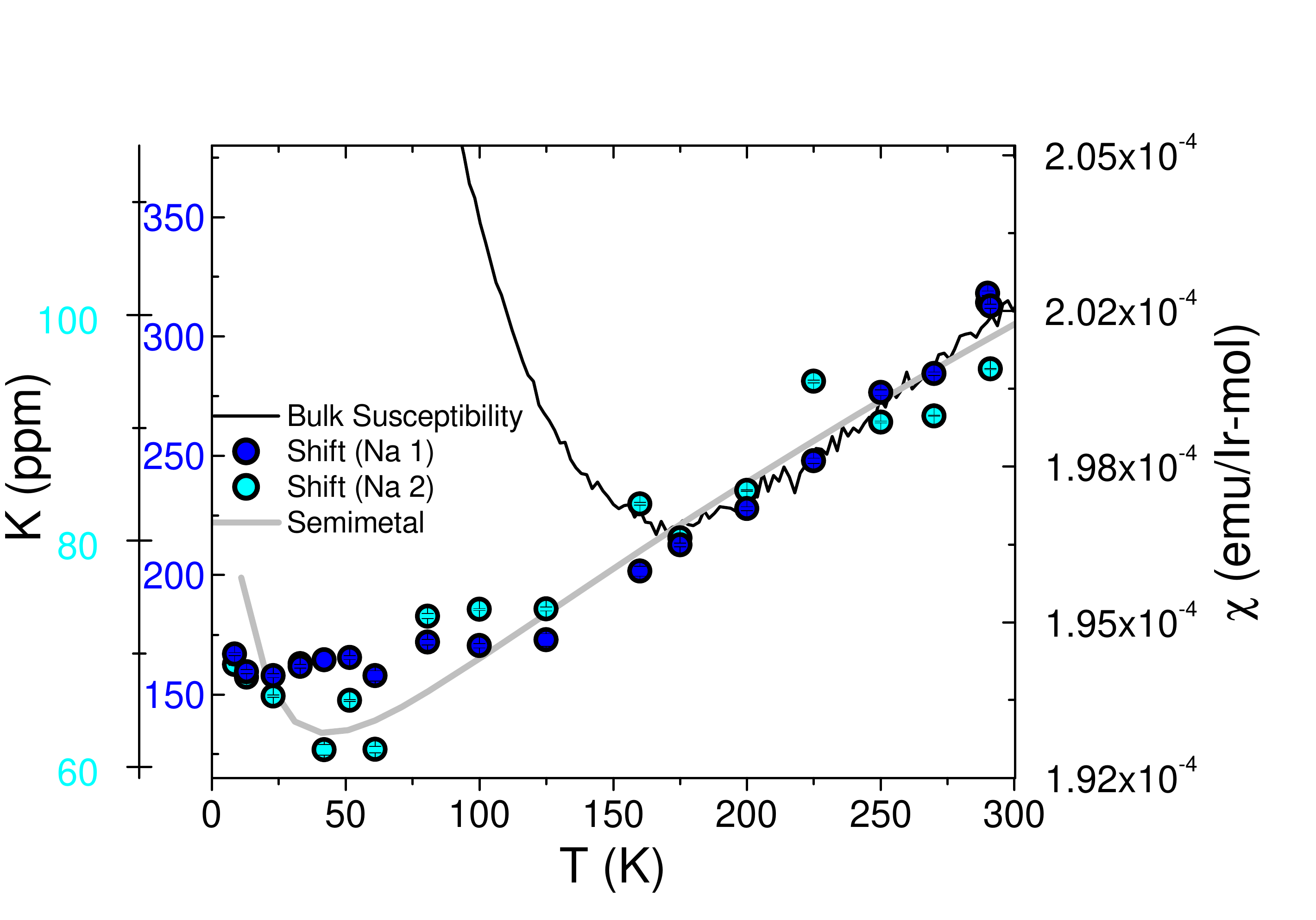}\\
\caption{\label{fig:shift}%
Macroscopic susceptibility \cite{Takayama2014} together with the shift measured at the two sodium sites. Three different scales are used to plot the different quantities, for the shift of the two peaks on the same scale, see the supplementary information. The shift tracks the susceptibility at high temperatures but at low temperatures, the behaviors for the two measurements diverge. The grey line represents a semimetal model following Ref.~\cite{Sales2010} as described in the text.
}
\end{figure}

The temperature dependence of the peak positions is shown in Fig.~\ref{fig:shift}. There is no reconstruction of the spectrum and the widths of the peaks remain the same throughout the studied temperature range meaning that no freezing takes place. As seen in Fig.~\ref{fig:shift} both peaks shift as a function of temperature at high temperatures, but have a much reduced temperature dependence below around 60 K. This yields the intrinsic susceptibility for this material. We note that the bulk susceptibility is not intrinsic, because the measurements are swamped by a Curie-like tail at low temperatures \cite{Takayama2014,Shockley2015,Sun2018}. This spurious contribution is equivalent to about 0.3 \% free spin 1/2 that we attribute to a tiny amount of an impurity phase.

\begin{figure}
\centering
\includegraphics[width={\columnwidth}]{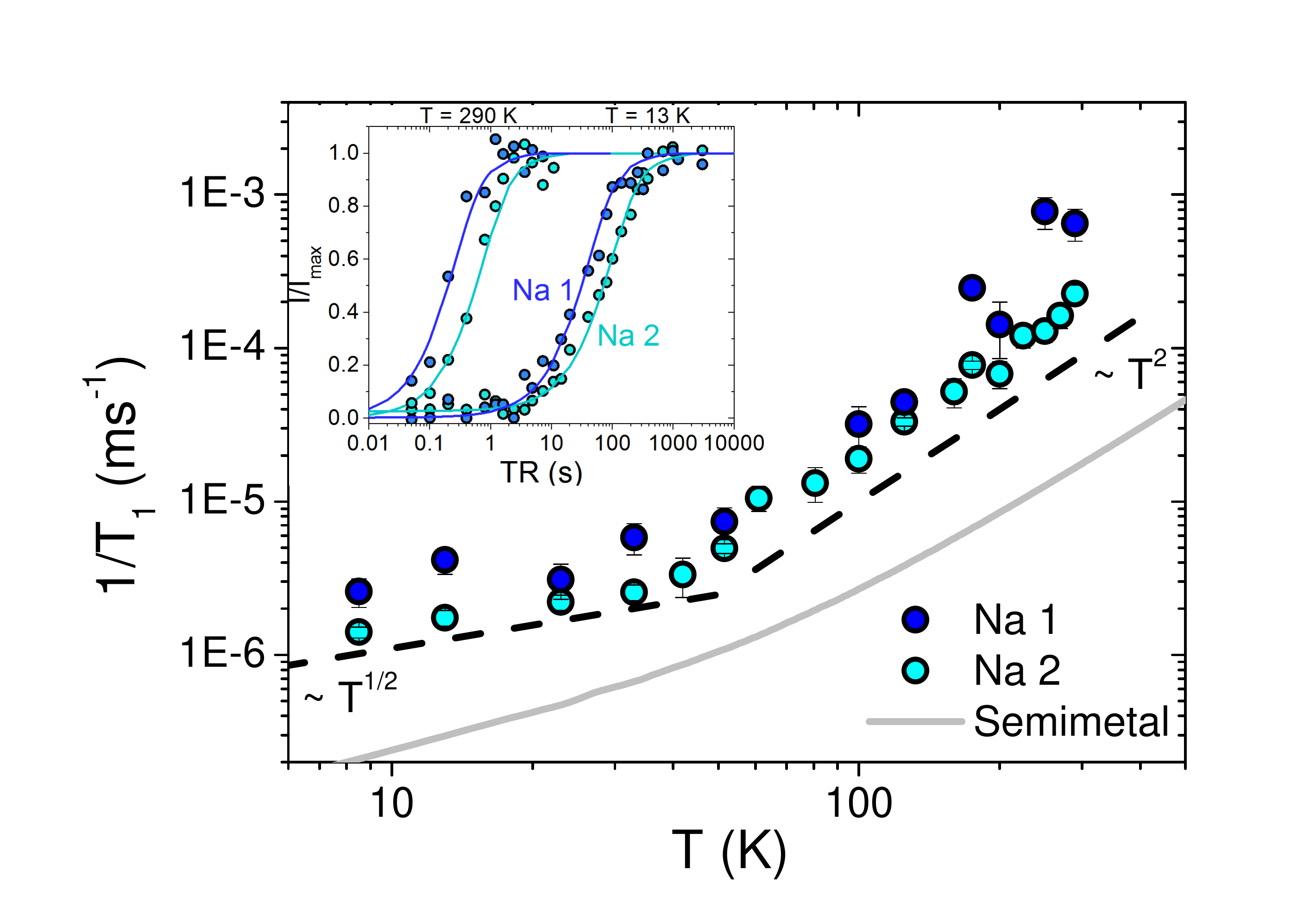}\\
\caption{\label{fig:Tone}%
Spin-lattice relaxation rate for the two sites as a function of temperature. Dashed lines are guides to the eye. The grey line is the temperature variation expected for a semimetal as explained in the text (the vertical position of this line is arbitrary). A characteristic recovery curve, measured at 290 K and 15 K on both Na sites is displayed in the inset.
}
\end{figure}

To access the dynamics of the system, relaxation experiments were performed. Series of nuclear magnetization recovery curves (inset of Fig.~\ref{fig:Tone}) were obtained at different temperatures to estimate the spin-lattice relaxation rate. The measurements were fitted to equation \ref{rec}, corresponding to the central line of I = 3/2 nucleus \cite{Suter1998}:

\begin{equation}\label{rec}
I (\mathrm{T_R})/ I_{Max}= 1-\frac{1}{10} e^{-(T_R/T_1)} -\frac{9}{10} e^{-(6T_R/T_1)}  \mathrm{,}
\end{equation}

\noindent where $I$ is the integrated intensity of the signal for the corresponding peak as a function of the repetition time $T_R$ and T$_1$ is the spin-lattice relaxation time. The data is well described using equation \ref{rec} without the need to introduce a distribution of the relaxation rates, in line with the absence of broadening of the spectrum. The inverse of the relaxation time, which corresponds to the momentum integrated dynamical structure factor at very low energy is shown in Fig.~\ref{fig:Tone} as a function of temperature. Interestingly, it also appears to exhibit different behavior depending on the temperature: at high temperatures the relaxation rate changes significantly faster than at low temperatures.

We now turn to the interpretation of our results. It is first important to acknowledge the unusual temperature dependence of the shift as a function of temperature. While in standard metals it is constant, in magnetic insulators at temperatures higher than the relevant interaction scales it decreases with temperature. Here the most probable reason for the observed temperature dependence of the susceptibility lies in the semimetal electronic structure, where the conduction and valence bands overlap in energy by E$_\mathrm{h}$. The unusual behavior of the susceptibility comes about from two effects. First, the chemical potential has a temperature dependence. Second, the density of states near the Fermi energy varies rapidly with energy. In particular, as the temperature is increased, the susceptibility is expected to increase due to an increase of the density of the carriers\cite{brouet2013}. Starting with a general expression of $\chi = - \mu_{B}^{2} \int \frac{df(E)}{dE}g(E) dE $, where $f(E)$ is the Fermi function, we follow the model worked out in the context of iron-based superconductors \cite{Sales2010}. We use the effective masses and the number of valleys determined by ab initio calculations \cite{Takayama2014}: 1.4, 2.3, 3.7 and 5.6 m$_0$ for hole bands and 1.8 m$_0$ for two electron bands. We can reproduce the intrinsic susceptibility data, by fitting only one parameter - the overlap of the conduction and valence bands, which we assume to be the same for all the bands and find to be E$_\mathrm{h}$ = 120(10) K. Note that the semimetal susceptibility, which is calculated without adjusting the scale, only accounts for the temperature-dependent part of the susceptibility. To reproduce the measured data, a temperature-independent contribution of 1.78*10$^{-4}$ emu/Ir-mol is added which is likely due to a Van Vleck-like contribution exceeding core diamagnetism, which comes in at -0.67*10$^{-4}$ emu/Ir-mol.

The trend is well-reproduced (Fig.~\ref{fig:shift}) but two points need to be addressed: first, at low temperatures there are deviations from the data, which could be due to an excess carrier concentration as well as due to an incomplete description of the band structure \cite{Sales2010}. Second, the obtained band overlap is smaller than the ab-initio estimate by a factor of 10, further suggesting a more complicated band structure. Correlations are likely the cause for such a reconstruction of the band structure, which are not accounted for by the ab-initio calculations in ref. \cite{Takayama2014}.

We now turn to the dynamics in this system. While the shift gives access to the uniform \textbf{q}$= 0$ susceptibility, the spin-lattice relaxation rate and its temperature dependence provide information on the nature and role of the spin excitations. As a starting point, we calculate the temperature dependence of the relaxation rate using the density of states $1/T_1 \propto \int g(E)^2 f(E) (1 - f(E)) dE $ and taking the modified $g(E)$ that was used to obtain the intrinsic susceptibility. As seen in Fig.~\ref{fig:Tone} this calculation captures the increase of the relaxation rate as a function of temperature, but the exact dependence is not reproduced.

The discrepancy is another indication of a more complex band structure and the role of correlations. The observed relaxation rate varies approximately as $1/T_1 \propto T^{1/2}$ at low temperatures, whereas the higher temperature region has $1/T_1 \propto T^2$ dependence. Interestingly in some Weyl semimetals such as TaP \cite{Yasuoka2017} and TaAs \cite{Wang2020}, a T$^{1/2}$ law was also found at low T and evolves to a stronger T$^3$ dependence at higher temperature. Clearly, the low density of states around the Fermi level is important. In the Weyl case, the density of states is vanishingly small at the touching point, whereas in a semimetal it remains finite which may produce a different power law at higher temperatures. While there is no satisfactory explanation for the T$^{1/2}$ low T behavior yet, it was suggested that the correlations among excited quasiparticles could be responsible \cite{Yasuoka2017}. Such a power law was also observed in a Mott insulator $\kappa$-(ET)$_2$Cu$_2$(CN)$_3$ and attributed to fluctuations which involve both the charge and spin \cite{Shimizu2006}. Since Na$_3$Ir$_3$O$_8$ is a semimetal on the verge of a transition to an insulator \cite{Takayama2014,Balodhi2015,Sun2018}, a similar scenario could be envisaged, especially at low temperatures, where the concentration of the carriers is low.

\begin{figure}
\centering
\includegraphics[width={\columnwidth}]{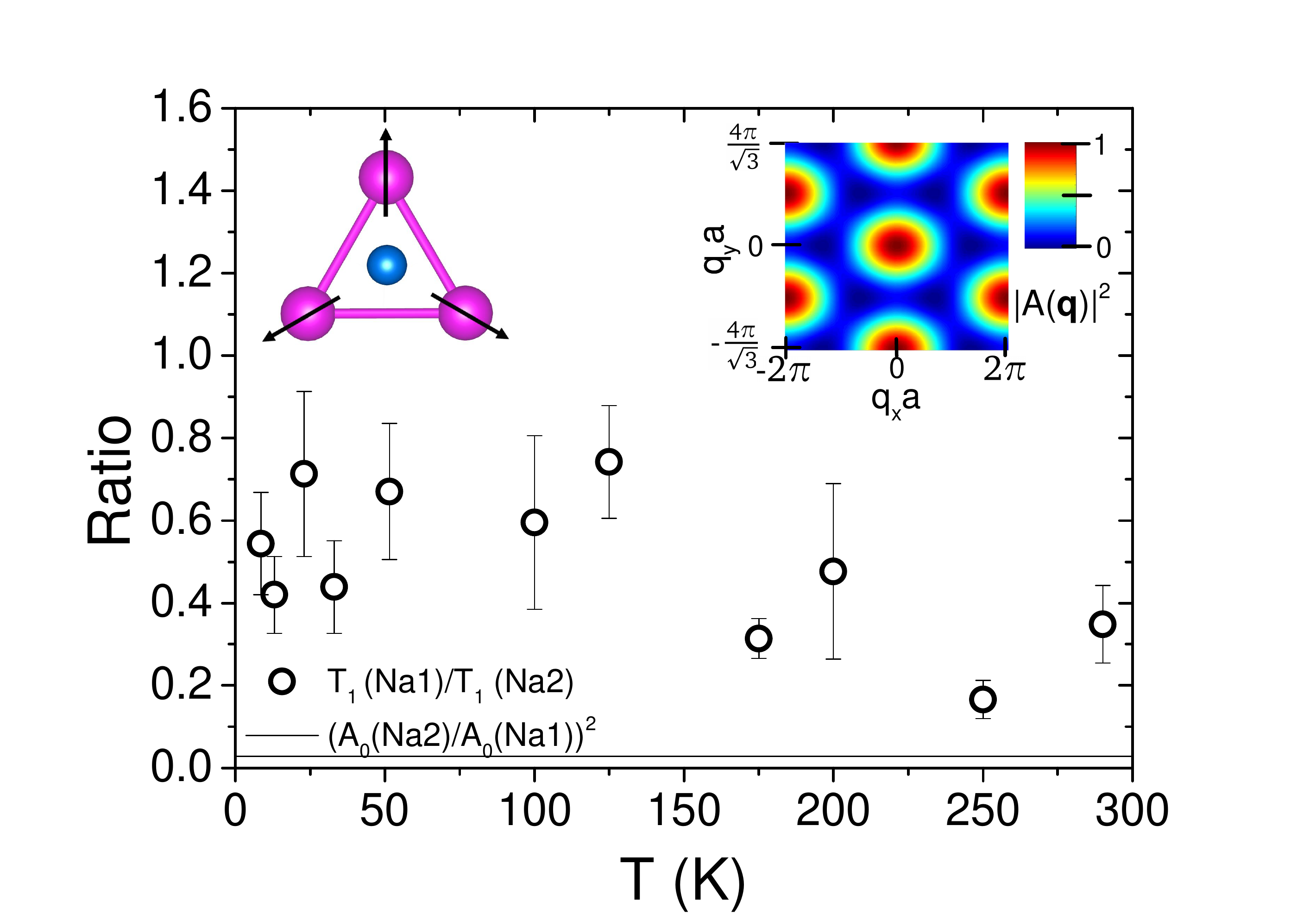}\\
\caption{\label{fig:A0rat}%
The ratio of T$_1$ values for two Na ions (symbols) compared to the square of the hyperfine coupling ratio (solid line). In a weakly correlated metal the two quantities are the same. The fact that they are not points to a q-dependent dynamic susceptibility. The top left inset shows the type of local correlations that will be filtered if a nucleus is positioned equidistantly within or above a triangle. A corresponding reciprocal space illustration is shown in the top-right inset, where the form factor for a fictitious triangular lattice is calculated and dark regions correspond to the 120$^o$ fluctuations that would be filtered.
}
\end{figure}

Leaving the exact temperature dependence of the relaxation aside, we turn to the robust observations of the differences between the relaxation rates of the two nuclei to study the nature of the correlations. Since there are two inequivalent Na sites in the system, one can use the ratio of the measured properties of the two sites to characterise the correlations at play. First, by inspecting the shift displayed in Fig.~\ref{fig:shift} in the range where there is a change, one finds that the ratio of the hyperfine coupling for the two sites is $\frac{\Delta K (\mathrm{Na2})}{\Delta K (\mathrm{Na1})}  = \frac{A_0(\mathrm{Na2})}{A_0(\mathrm{Na1})} \approx \frac{1}{6}$. Essentially the same value is also obtained by analyzing the shift and bulk susceptibility using the Clogston-Jaccarino plot (see \cite{SI2020}). As the relaxation rate depends on the square of the hyperfine coupling and on the number of nearest neighbors, the difference in the relaxation rates in the absence of correlations is expected to be of the order of $\approx$ 1/54 - much smaller than what is observed. The ratio itself is plotted for clarity in Fig.~\ref{fig:A0rat}. This again indicates that the weak-correlation approach is insufficient to explain the behavior in this system.

A likely reason for the discrepancy in the relaxation rates for the two different nuclei is the presence of antiferromagnetic correlations revealed by the virtue of the different local environment for the two sites. As can be seen in Fig.~\ref{fig:HKGM}, the Na 1 nucleus is located in between two iridium triangles. In the case of antiferromagnetic correlations favoring the 120$^\circ$ motif, the fluctuations at the $^{23}$Na nucleus will be screened (left inset of Fig.~\ref{fig:A0rat}). The Na 2 nucleus appears not to be located in such a symmetric position: it is linked to one triangle, which will be filtered as for Na 1, but the other six Ir ions are from three different triangles. This can be formalized for both sites by considering the general expression for the relaxation rate $\frac{1}{T_1}= \propto \sum_{\bm{q}} |A (\bm{q})|^2 \chi''(\bm{q},0) $, where $\chi''(\bm{q},0)$ is the imaginary part of the dynamical spin susceptibility and $A (\bm{q})$ is the hyperfine form factor, which encapsulates the sensitivity of the fluctuations to different wavevectors. It depends on the geometrical arrangement of the probing nucleus and the magnetic ions. The expression in the reciprocal space is the sum of the contributions from the nearest Ir neighbors at a distance r$_i$: $A (\bm{q})= \sum_i A_i \exp(-i \bm{q r_i})$. The A$_i$ is in principle different for every Ir, but in the case of Na 1, all Ir atoms have the same distance and linkage to the Na 1 nucleus. First we calculate the hyperfine form factor giving the \textbf{q} dependence of the  A(\textbf{q}) in simplified coordinates using one triangle with a lattice constant a. This is displayed in the right inset of Fig.~\ref{fig:A0rat} and corresponds locally to the exact situation for the Na 1 nucleus. Dark regions in the form factor show the modes that would be filtered, which correspond to the 120$^o$ motif. This explains the deviation from the expected ratio of the relaxation rates - these local modes are filtered by the Na 1 nucleus.

We now ask further, whether there are long-range modes that satisfy such restrictions and turn to the comparison between the Na 1 and the Na 2. Quite unexpectedly we find that simple modes based on the 120$^o$ motif with \textit{long range} coherence will be filtered both for Na 1 and Na 2 \cite{SI2020}. In terms of form factor, this means that only very unusual locations of the q-space have strong difference in filtering long range modes. While some exotic dynamics cannot be strictly excluded, the most likely reason for strong filtering in the case of Na 1 and not in the Na 2 is the short-range character of the correlations. The 120$^o$ dynamic correlations are restricted to approximately one triangle only.

To summarize, our measurements position the frustrated hyperkagome compound Na$_3$Ir$_3$O$_8$ as a semimetal with short range magnetic correlations, reminiscent of the spin liquid parent compound.

We hope that our study will stimulate theoretical efforts to understand the effect of doping a spin liquid in the case of iridates. Experimentally, obtaining the bandstructure would be desirable as well as probing the magnetic correlations further by diffuse scattering. Finally, in an attempt to dope the system further and stabilize other non-conventional states, gating the material would be illuminating.

This project is supported by Agence Nationale de la Recherche under SOCRATE (ANR-15-CE30-0009-01) and LINK (ANR-18-CE30-0022) projects. The work of GS is funded by the Swiss National Science Foundation Mobility grant P2EZP2-178604 and PALM LabEx grant ANR-10-LABX-0039-PALM.

\bibliography{Na3Ir3O8bib}

\end{document}